\def\be{\begin{equation}}
\def\ee{\end{equation}}

\documentclass[12pt,a4paper,epsfig]{iopart}
\usepackage{epsfig}

\usepackage{color}

\usepackage{graphicx}% Include figure files
\usepackage{dcolumn}% Align table columns on decimal point
\usepackage{bm}% bold math

\begin{document}

\title[Modeling a Schottky-barrier carbon nanotube field-effect
transistor] {Modeling a Schottky-barrier carbon nanotube
field-effect transistor with ferromagnetic contacts}

\author{S. Krompiewski}
\address{Institute of Molecular Physics, Polish Academy of
Sciences, M. Smoluchowskiego 17, 60-179 Pozna\'n, Poland}

\date{\today}

\begin{abstract}
  In this study, a model of a Schottky-barrier carbon nanotube field-effect transistor (CNT-FET),
  with ferromagnetic contacts, has been developed.
  The emphasis is put on analysis of current-voltage characteristics as well
  as shot (and thermal) noise. The method is based on the tight-binding model
  and the non-equilibrium Green's function technique.
  The calculations show that, at room temperature, the shot noise of the CNT FET
  is Poissonian in the sub-threshold region,
  whereas in elevated gate and drain/source voltage regions the Fano factor gets strongly reduced.
  Moreover, transport properties strongly depend on relative magnetization orientations in the
  source and drain contacts. In particular, one observes quite a large tunnel magnetoresistance,
  whose absolute value may exceed 50\%.
\end{abstract}

\pacs{85.75.Hh, 75.47.De, 42.50.Lc, 73.63.Fg}
% FET, GMR, quantum noise, nanotubes el. transport

\maketitle

\section{Introduction}

In view of  well-known size reduction problems of the conventional
Si-based electronics, there have recently been intensive studies
on new technologies based on nanostructured materials which are
formed by organized growth and self-assembly methods
\cite{hoenlein}.
%self-organized bottom-up approaches.
A remarkable example of such self-organized structures are carbon
nanotubes, which due to their fascinating physical properties have
been studied for more than one and a half decade now and are
believed to have commercial applications in the near future. In
particular a nanotube-based transistor was first demonstrated in
\cite{tans98}, and has since been studied extensively
(\cite{leonard99}-\cite{javey02}). Carbon nanotubes are also very
promising for spintronic applications, they may act as
magneto-resistive switches \cite{coskun,zaric}, and can maintain
spin coherence over long distances. The latter is crucial for the
tunnel (giant) magnetoresistance TMR (GMR) effect to be
observable. In fact the TMR effect in CNTs was first reported in
\cite{tsukagoshi99}, where the spin diffusion length of electrons
flowing through a carbon tube was estimated to be on the order of
routinely used contact separations in electric transport
measurements. Subsequent studies confirm that the TMR effect in
CNTs can be quite large, ranging typically from a few up to
several tens percent (\cite{mehrez}-\cite{hueso}). Here it is
shown that also for the Schottky-barrier nanotube transistor the
situation is likewise.

Shot noise is another important phenomenon of present interest. It
only appears in a non-equilibrium situation (finite source/drain
voltage) and originates from the discreteness of the electron
charge. In spite of there being a lot of papers on shot noise in
nanostructures (\cite{blanter}-\cite{wu07}) (see \cite{blanter05}
for a comprehensive review of recent advances), there have
hitherto been no attempts to investigate shot noise in
ferromagnetically contacted Schottky-barrier nanotube transistors
(to the author's knowledge).

The paper is organized in the following way: In Sec.~2 a short
discussion of the adopted approach is presented, including the way
the electrostatics has been dealt with, and basic equations
concerning the non-equilibrium Grean's function method. Section 3
is devoted to main results of the paper, i.e. current and shot
noise dependences on the gate and drain/source voltages, as well
as the impact of ferromagnetic electrodes on the above mentioned
transport characteristics. Finally, Sec.~4 concludes the papers.

\section{Methodology}

A nanotransistor under consideration consists of a single wall
carbon nanotube (CNT), described in terms of $\pi$-orbital
electrons, end-contacted to s-type itinerant-electron slabs of fcc
(111) crystallographical structure. The latter may be either
paramagnetic or ferromagnetic. The system considered here is
end-contacted, in contrast to other wide-spread geometries, like
those of side-contacted or embedded ones. The device has been
relaxed in order to find energetically favorable positions of
interface atoms represented by big and small spheres with
diameters of $d_M = 2.51 \AA$ and $d_C = 1.421 \AA$, for metal and
carbon, respectively \cite{krompiewski06,ustron}. The studies are
carried out within the framework of the tight-binding model and
the non-equilibrium Green's function technique. The basic idea of
this study is to adopt the methodology known for one-dimensional
transistor models \cite{indlekofer05,pikus,auth97}, so as to make
it useful for description of a Schottky-barrier nanotube
transistor. This method bears much similarity to earlier methods
\cite{odintsov,nakanishi,anantram06}, except that here the
calculations are performed in real space for CNT of finite length,
and there is no need of using any ideal nanotube energy spectra
(nor the Wentzel-Kramers-Brilloin, WBK, approximation). It should
be however stated in this context that there is a great deal of
theoretical studies on nanotube transistors (see e.g
\cite{guo}-\cite{leonard}), which successfully avoid reducing the
Poisson equation to the 1-D problem, and handle thereby the
electrostatics on a higher level than it is done here.
Nevertheless, the present approach, while benefiting enormously
from the mathematical simplicity, still leads to qualitatively
correct results. The one-dimensional Poisson equation and its
solutions are regarded here as spin-dependent(however spin indexes
are skipped for brevity), and read \cite{indlekofer05}

\begin{equation} \label{Poisson}
 \frac{\partial^{2}}{\partial x^2}V(x)+\frac{1}{\lambda^2} \left[
V_G-V(x) \right] +\frac{1}{\epsilon_0 \epsilon_{CNT}A}\rho(x)= 0,
\end{equation}

\begin{equation}\label{electrostatics}
V(x) = \frac{1}{\epsilon_0 \epsilon_{CNT} A} \int dx' v(x,x')
\rho(x')+V_{ext}(x),
\end{equation}

\begin{eqnarray}
\label{1} v(x,x') = \frac{\lambda}{2} [
e^{-\left|x-x'\right|/\lambda}-e^{-(x+x')/\lambda} + \hskip1cm
\nonumber \\
e^{-L/\lambda} \left( \cosh\frac{x-x'}{\lambda}-
\cosh\frac{x+x'}{\lambda} \right) /\sinh\frac{L}{\lambda}], \;
\hskip0.3cm
\end{eqnarray}

\begin{eqnarray}
\label{2} V_{ext}(x)& = & [V_S \sinh\frac{L-x}{\lambda}+V_D
\sinh\frac{x}{\lambda}]/\sinh\frac{L}{\lambda}  \hskip1cm
\nonumber \\
& + & \frac{1}{\lambda^2}  \int dx' v(x,x')V_G, \;
%\hskip0.3cm
\end{eqnarray}

where $ \lambda=(R/2)\sqrt{2 (\epsilon_{CNT}/\epsilon_{ox}) \ln(1+
d_{ox}/R)+ 1 }$ is the effective screening length \cite{auth97},
$\epsilon_{CNT}$ and $R$ stand for the dielectric constant and the
radius of the CNT, whereas the subscript $ox$ refers to the
coaxial gate oxide layer $SiO_2$ ($d_{ox}$ in thickness). The
other symbols have the following meaning: $V_S$, $V_D$, $V_G$ -
source, drain and gate voltages; $L$ - the length of the CNT;
$f_{S,D}$ - Fermi function at $V_S$ and $V_D$; $A$ - the CNT
cross-sectional area.

As regards the non-equilibrium Green's function (GF) technique, a
recursive method, similar to that of \cite{SK02} was used, with
obvious modifications consisting in including the potential
profile V(x) (Eq.~\ref{electrostatics}) to the CNT Hamiltonian,
and the $V_S$ and $V_D$ voltages in the contacts. This procedure
corresponds to the Hartree approximation, which is sufficient for
the present room-temperature model, far beyond the Coulomb
blockade and Kondo regimes. With this proviso one can write down
equations for spin-projected current (I), zero-frequency noise
power (S) and transmission matrix (T) as follows

\begin{eqnarray}\label{IST}
I & = & \frac{e}{h} \int{dE \, (f_S - f_D) \,  \,Tr \,[T(E)]}, \nonumber\\[3mm]
S & = & \frac{2e^2}{h} \int dE \, \, \{[ f_S(1-f_S) + f_D(1-f_D)] \, Tr[T(E)]  \nonumber\\
% [2mm]
& + &  (f_S-f_D)^2 \, Tr[T(E) \left(1-T(E) \right)] \}, \nonumber\\[3mm]
T & = & \Gamma_S G^r \Gamma_D G^a,
\end{eqnarray}

where trace ($Tr$) is taken over the orbital indexes, and $
\Gamma_\alpha = i(\Sigma^r_\alpha - \Sigma^a_\alpha) $
 (with $ \alpha= $ S, D for source and drain, respectively). The first term of the noise power S is the thermal noise,
 it disappears at zero temperature, but dominates at finite temperatures and vanishing
source/drain potentials. The Fano factor is defined here -
assuming that both spin channels are independent (no spin
relaxation processes) - in a usual way as
$F=(S^\uparrow+S^\downarrow)/(2e(I^\uparrow+I^\downarrow))$, so
that it is one for the Schottky shot noise (Poissonian limit).

Instead of a rather common wide-band type approximation, the
present method uses energy-dependent self-energies
($\Sigma_\alpha(E)$) which have been expressed in terms of:
\emph{(i)} recursively computed surface Green functions
($g_\alpha(E)$) of infinite fcc-(111) contacts, and \emph{(ii)}
the CNT/contact coupling matrices ($v_c$), i.e
$\Sigma_\alpha(E)=v_c \, g_\alpha(E) \, v_c^\dag$. Here the
retarded, $G^r$, (advanced, $G^a$) GF is a matrix of rank equal to
the number of atoms in the unit cell it refers to. The density
matrix is defined as $\hat \rho=-i/(2 \pi) \int dE G^<(E)$ in
terms of the lesser GF, which can be brought into the following
form:
%\begin{equation} \label{less}
$G^< = -f_D (G^r-G^a)+i(f_S-f_D)G^r \Gamma_S G^a.$
%\end{equation}
 While integrating $G^<$, its first
 term contribution is found by integral over the contour in the complex energy
 plane, and the second one - by integrating along the real energy axis (using the
 Gaussian quadrature). The so-determined integral makes it possible to get an extra (excess or
 deficit) electric charge per CNT atom, with respect to the charge neutrality situation
 (no external potentials). The procedure
 is self-consistent because the potential $V$ enters the diagonal of the
 Hamiltonian, which in turn, is present in the GF denominator. In order to make the recursive algorithm
 effectively work, and speed up
 the convergence, an additional assumption has been made, \emph{viz.} the potential
 profile is only dependent on the unit cell number and remains constant within the given unit cell.

\section{Results and discussion}

\begin{figure}[t]
\hspace*{-0.5cm}
\includegraphics[scale=0.6]{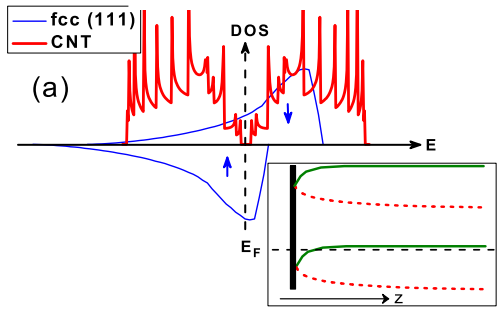}
\vspace*{-0.5cm} \caption{ \label{schematic} Spin-split surface
densities of states (DOS) of a contact (thin line). In the
antiparallel configuration the second contact has got interchanged
$\uparrow,\downarrow$ sub-bands (no interchange in the parallel
case). The thick curve is for the CNT DOS.
 \emph{Inset}: Band bending at the interface for $V_G>0$ (short-dash line)
 and $V_G<0$ (solid line).}
\end{figure}

Detailed computations have been carried out for a zigzag
semiconducting CNT with a chiral vector (n, 0) and n=14. The
radius $R=a \, \, n/(2 \pi)$, and the energy gap $\Delta=2 \pi
t/(\sqrt{3}n)$, where a=0.249 nm is the graphene lattice constant
and $t=2.7$ eV is the $\pi$-orbital hopping integral.
Additionally, it has been set: $\epsilon_{CNT}=1 $,
$\epsilon_{ox}= 4$, $d_{ox}=2.5$ nm. The CNT is $l$ unit cells
long ($L=l \, a \, \sqrt{3} $), and the calculations are performed
for $l=100$ (as well as for the ultra-short length $l=15$). It is
well known that the Schottky-barrier at a semiconductor/metal
interface depends mainly on relative work functions (WF) of
materials which form the junction \cite{chen,leonard,nosho}. Here
it is assumed that the metal WF is greater than that of the CNT,
as e.g. for Pt, Au, Ni, Fe (as opposite to Ti), and locate the
Fermi energies below the mid-gap. To be specific, the metal Fermi
energy has been fixed at one-quarter of the band gap. On- and
off-states of the FET are illustrated in Inset to
Fig.~\ref{schematic}, where the lower curves correspond to the
maximum of the valence band and the upper curves correspond to the
minimum of the conduction band.
Accordingly, the present model-device represents an "$E_F$ close
to valence band maximum" case (cf.~\cite{heinze02}).
 It is seen that, depending on
the gate-voltage, the CNT-FET is either in the on-state (solid
lines) or the off-state (short-dash lines). The former is due to
the hole tunnelling through the Schottky-barrier, whereas the
latter results from the fact that the transport energy-window
falls within the energy gap. The location of $E_F$ beyond the
mid-gap of the isolated CNT introduces the electron-hole
asymmetry, as shown recently \cite{wang} a similar effect may be
caused by the presence of the inner uncontacted shell in a
double-walled CNT, leading essentially to similar results for
current-voltage characteristics (in the non-magnetic case).
\begin{figure}[h]
\hspace*{-0.5cm}
\includegraphics[scale=0.6]{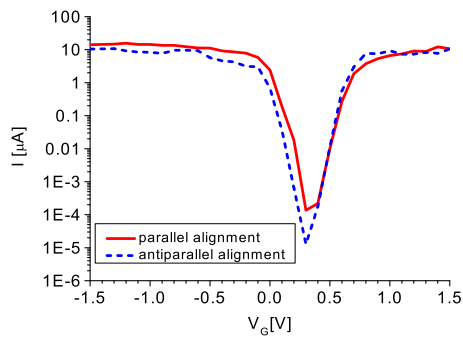}
 \vspace*{-0.5cm} \caption{ \label{F1a} The current \emph{vs.}
 gate-voltage characteristics for the parallel alignment of the contact magnetizations
 (solid
line) and the antiparallel alignment (dashed line). The
drain/source voltage is equal to 0.2V, length L=100 unit cells,
and temperature T$=300^\circ $K.}
\end{figure}
Figure \ref{F1a} clearly shows that depending on boundary
conditions imposed by the magnetic contacts [parallel alignment
(\emph{P}) vs. antiparallel alignment (\emph{AP})] the
current-gate voltage characteristics differ considerably from each
other. Notably, the ON/OFF current ratio is also different for
both the alignments. This is because the sub-threshold region
reveals a conventional (Julliere's type \cite{julliere}) behavior
with $I_P$ clearly greater than $I_{AP}$, whereas beyond this
region both the currents differ far less from each other. Within
the "non-Julliere's" region the negative TMR apparently appears as
a result of the shift of threshold $V_G$ for electron conduction
between P and AP configurations.
 Figure \ref{F3} presents the corresponding
Fano factors. The shot noise of the CNT FET is Poissonian (F=1) in
the sub-threshold region, and gets substantially reduced for
elevated $V_G$ and $V_{DS}$. A still stronger reduction would be
possible if the transport regime e.g. was either ballistic or of
electron billiard type (with a short dwell time)
\cite{blanter,BS}. As expected, magnetic conditions of the
contacts, which strongly influence the current, have also a
considerable effect on the noise. It is readily seen from
Fig.~\ref{F3} that the Fano factor corresponding to the P
alignment, when compared to that of the AP alignment, is
predominantly suppressed. The suppression takes place in a large
region of the gate voltage, where current in the P configuration
is greater than in the AP configuration (cf. Fig.~\ref{F1a}), in
concord with the fact that $F \sim 1/I$. It should be stressed
that the transport regime considered here is phase coherent,
although with noticeably reduced transmission due to the presence
of Schottky-barriers. That is why the length effects, as shown in
the Inset to Fig.3, are quite moderate, in contrast to what could
be expected in the diffusive transport regime. Spin-dependent
electron wave interferences are responsible for non-monotonic
behavior of presented plots and accompanying sharp features.
Similar features have been reported both in conductance and TMR
\cite{mehrez,cottet06,man} as well as in the differential Fano
factor \cite{wu07}.
\begin{figure}[t]
\hspace*{-0.5cm}
\includegraphics[scale=0.60]{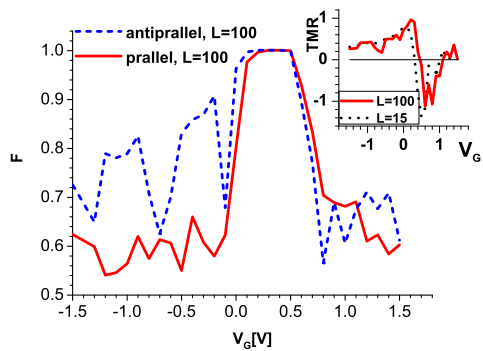}
 \vspace*{-0.5cm} \caption{ \label{F3}
 As in Fig.~\ref{F1a}, but for the Fano factor. Note that reversal of the magnetic configuration
 from AP to P, for $V_G<0$, may result in a considerable suppression of F. \emph{Inset}:
 TMR for 2 different CNT lengths and $V_{DS}=0.2V$. }
\end{figure}
The TMR is defined here in a "pessimistic" way as
   $ TMR = (I_{\rm P} - I_{\rm AP})/I_{\rm P} $,
where $I_{\rm P}$ ($I_{\rm AP}$) is the current
\begin{figure}[h]
\hspace*{-0.5cm}
\includegraphics[scale=0.60]{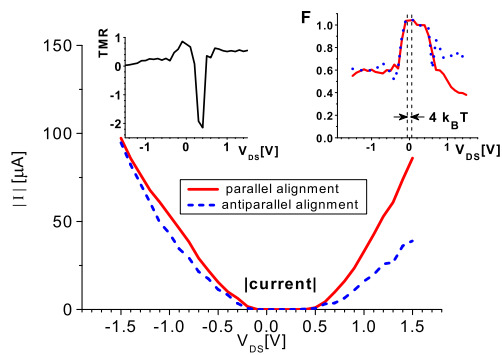}
 \vspace*{-0.5cm} \caption{ \label{F2}
 Drain/source voltage dependence of I, TMR and F (main, left and right panels) for
 $V_{G}=0.3 V$, T=300K and L=15. In the interval $\pm 2k_B T$ around $V_{DS}=0$ thermal noise
 dominates and $I \rightarrow 0$, so F is ill-defined.}
\end{figure}
flowing through the system in the P (AP) magnetic configuration.
The calculations are performed for spin polarizations in the
ferromagnetic contacts equal to
$(n_\uparrow-n_\downarrow)/(n_\uparrow+n_\downarrow)=0.5$
($n_\sigma$ is the number of $\sigma$-spin electrons per atom).
The results concerning TMR against $V_G$ and $V_{DS}$ are
presented in Figs. 3 and 4, respectively. The corresponding Insets
clearly show that the TMR absolute value may be quite substantial
and exceed several tens percent both in the sub-threshold region
and at elevated voltages. Interestingly enough, TMR can assume
large negative values (inverse TMR) and is strongly sensitive to
both the gate- and $V_{DS}$-voltages. Another important
observation is a striking asymmetry of the Fano factor and the GMR
effect with respect to the sign of the voltage ($p$-channel vs.
$n$-channel). These features are due to the aforementioned
location of $E_F$ with respect to the mid-gap, which makes the
Schottky barrier for electrons higher than that for holes, and
consequently the hole current (for negative voltages) is higher
than the electron current (for positive voltages). It is
noteworthy, that the oscillations present in
Figs.~\ref{F1a}-\ref{F2} are due to the size-dependent
quantization of the CNT energy-levels, so the local maxima
(minima) therein depend on to which extent the voltage-driven
active channels and the transport energy-window fit together. This
mechanism is also responsible for the occurrence of the inverse
GMR, in fact it is similar to the mechanism based on resonant
tunneling suggested in \cite{sch06}.

\section{Conclusions}

An approach has been developed to describe nanotube transistors,
which combines a simplified (analytical) treatment of the
electrostatics, and the state-of-the-art non-equilibrium GF
technique. Unprecedented studies of electric transport through
Schottky-barrier CNT transistor with ferromagnetic electrodes have
been carried out. It has been shown that the tunnel ability of
particles depends strongly on the alignment of contact
magnetizations. Remarkably, the tunnel magnetoresistance can
exceed 50\%, whereas the Fano factor is usually subjected to some
extra suppression when the magnetization alignment gets reoriented
from the antiparallel to the parallel one, improving thereby the
noise-to-signal ratio. These two findings are relevant for
prospective applications of nanotubes in spintronics.

%\vspace{-1cm}
\begin{ack}
This work was suported by the EU FP6 grants: CARDEQ under contract
No. IST-021285-2, and SPINTRA under contract No.
ERAS-CT-2003-980409.
\end{ack}

\end{document}